\documentclass[pdflatex,sn-mathphys-num]{sn-jnl}

\usepackage{graphicx}%
\usepackage{multirow}%
\usepackage{amsmath,amssymb,amsfonts}%
\usepackage{amsthm}%
\usepackage{mathrsfs}%
\usepackage[title]{appendix}%
\usepackage{xcolor}%
\usepackage{textcomp}%
\usepackage{manyfoot}%
\usepackage{booktabs}%
\usepackage{algorithm}%
\usepackage{algorithmicx}%
\usepackage{algpseudocode}%
\usepackage{listings}%
\usepackage{bibunits}
\usepackage{hyperref}
\usepackage{acronym}%
\usepackage{tikz}%
\usepackage{pgfplots}%
\pgfplotsset{compat=1.18}%
\usepgfplotslibrary{groupplots}%
\usetikzlibrary{arrows,automata}
\usetikzlibrary{shadings}
\usetikzlibrary{spy}%
\usepackage[font=footnotesize]{caption}
\usepackage[font=footnotesize,labelformat=simple]{subcaption} 

\usepackage{tcolorbox}
\tcbuselibrary{breakable}
\usepackage{pdfpages}

\acrodef{NPI}	    [NPI]	    {Non Pharmaceutical Intervention}
\acrodef{SIR}	    [SIR]	    {Susceptible Infected Removed}
\acrodef{SEIR}	    [SEIR]	    {Susceptible Exposed Infected Removed}
\acrodef{SI}	    [SI]	    {Susceptible Infected}
\acrodef{MDP}	    [MDP]	    {Markov Decision Process}
\acrodef{MJP}	    [MJP]	    {Markov Jump Process}
\acrodefplural{MDP}  [MDPs]	    {Markov Decision Processes}
\acrodef{VI}	    [VI]          {Value Iteration}

\DeclareMathOperator*{\argmin}{argmin}

\definecolor{bckgrddblue}{RGB}{174,197,230}
\colorlet{lightblue}{bckgrddblue!25!white}
\colorlet{innercircle}{bckgrddblue!10!white}           %
\colorlet{externalcircle}{rgb,255:red,21; green,45; blue,80}
\definecolor{myorange}{RGB}{214,118,48}%
\definecolor{myinfectcolor}{RGB}{40,42,115}%
\definecolor{myrecoverycolor}{RGB}{127,20,22}%

\newtcolorbox[auto counter]{pabox}[2][]{colback=green!5!white,colframe=green!70!black,fonttitle=\bfseries,title=Theory box~\thetcbcounter #2,#1}
\newtcolorbox[auto counter]{mabox}[2][]{colback=lightblue,colframe={rgb,255:red,21; green,45; blue,80},fonttitle=\bfseries,title=Epidemic models box~\thetcbcounter #2,#1}

\theoremstyle{thmstyleone}%
\theoremstyle{thmstyletwo}%

\theoremstyle{thmstylethree}%

\begin{document}

\title[Fundamental limits on taming infectious disease epidemics]{Fundamental limits on taming infectious disease epidemics}

\author[1]{\fnm{Giovanni} \sur{Pugliese Carratelli}}\email{gp459@cam.ac.uk}

\author[2]{\fnm{Xiaodong} \sur{Cheng}}\email{xiaodong.cheng@wur.nl}

\author[3]{\fnm{Kris V.} \sur{Parag}}\email{k.parag@imperial.ac.uk}

\author[1]{\fnm{Ioannis} \sur{Lestas}}\email{icl20@cam.ac.uk}

\affil[1]{\orgdiv{Department of Engineering}, \orgname{University of Cambridge}, \orgaddress{\street{Trumpington street}, \city{Cambridge}, \postcode{CB1 PZ},  \country{United Kingdom}}}

\affil[2]{\orgdiv{Biometris group}, \orgname{Wageningen University}, \orgaddress{\street{Radix building 107, Droevendaalsesteeg 1}, \city{Wageningen}, \postcode{6708 PB},  \country{The Netherlands}}}

\affil[3]{\orgdiv{MRC Centre for Global Infectious Disease Analysis}, \orgname{Imperial College London}, \orgaddress{\street{90 Wood Lane}, \city{London}, \postcode{W12 0BZ},  \country{United Kingdom}}}

\begin{bibunit}[naturemag]

\abstract{Epidemic control frequently relies on adjusting interventions based on prevalence \cite{Kraemer2025,Berger2021,Tsay2020}.  But designing such policies is a highly non-trivial problem \cite{Dietz2000,Bedford2019} due to uncertain intervention effects \cite{Flaxman2020}, costs \cite{Haw2022,Tenreyro,Coibion2020} and the difficulty of quantifying key transmission mechanisms and parameters \cite{PastorSatorras2001,Ferguson2007,PastorSatorras2015,ChangeofPars}. Here, using exact mathematical and computational methods, we reveal a fundamental limit in epidemic control in that prevalence feedback policies are outperformed by a single optimally chosen constant control level. Specifically, we find no incentive to use prevalence based control under a wide class of cost functions that depend arbitrarily on interventions and scale with infections. We also identify regimes where prevalence feedback is beneficial. Our results challenge the current understanding that prevalence based interventions are required for epidemic control and suggest that, for many classes of epidemics, interventions should not be varied unless the epidemic is near the herd immunity threshold.}

\maketitle
Effectively controlling emerging infectious diseases is a complex problem because of the inevitable trade-offs between the impact of interventions and their effects on society. Decisions made at the current time also impact the future course of the epidemic, adding further complexity to the decision-making process. The course of the epidemic is difficult to model \cite{Giordano2020,Zino2021,Nowzari2016}, requires high quality data \cite{Vespignani2020}, and high uncertainty surrounds the complex transmission mechanism \cite{Britton2015,Keeling2007,Wearing2005}. Moreover, the various costs that need to be considered \cite{Chater2020,Barnett2023} and the effectiveness of the interventions on the spread are challenging to quantify \cite{Flaxman2020}.

Mitigation strategies are in practice frequently chosen to be to some degree sensitive to infections \cite{Ferguson2005,Ferguson2006,Longini2005,DiLauro2021}, \emph{i.e.} a form of negative feedback such that when infections rise, increasingly restrictive interventions are implemented and when observed infections decrease interventions are eased. Effective policy design principles that facilitate policy recommendation are however difficult to establish \cite{Morris2021} as the considered costs can lead to different policies. For example, restrictive measures are used for high hospitalisation costs and low intervention costs while less restrictive interventions can be optimal for low hospitalisation costs and case to fatality ratios.
The complex intertwine of costs and epidemic models raises therefore a central concern of which policies are effective and to what extent epidemics can be controlled. 

Here we show using exact methods from stochastic control that for a wide class of costs and realistic epidemics constant interventions are not only simpler, but also optimal---outperforming strategies that adapt to prevalence.

\begin{figure}[H]
	\centering
	\includegraphics[width=0.53\columnwidth]{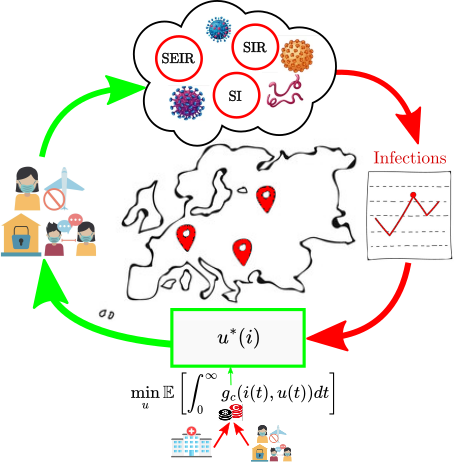}
\caption{\textbf{Schematic of optimal control for epidemics.} Finding an optimal mitigation policy is a complex task that involves optimising at any time $t$ over over all interventions and their effect over the entire future time horizon (see \eqref{eq.GenericCosts}). The costs and the epidemic parameters are also inevitably uncertain.
We investigate limitations of effective control for a broad set of the costs $g_c(i,u)$. We consider randomised versions of the \ac{SEIR}, \ac{SIR} and \ac{SI} specified in Epidemic models box \ref{modelbox}.}\label{fig.MainDiagram}
\end{figure}

In order to approach this problem we consider the problem of finding transmission mitigation measures $u$ that minimise the social/economic costs $g_c(i,u)$ associated with the number of infected $i$ and the mitigation measures, over an indefinite amount of time as in \eqref{eq.GenericCosts} (see Fig. \ref{fig.MainDiagram}).
We consider the epidemics to evolve probabilistically capturing the fact that infections occur as randomised events with prescribed rates as described in the Epidemics models box \ref{modelbox}. These models are more accurate than corresponding deterministic models and also incorporate the fact that epidemics are controlled by modulating the rates rather than individual events. Such a stochastic formulation further complicates the choice of mitigation measures $u$ as a multitude of different outcomes have to be considered. 

\begin{mabox}[label={modelbox},breakable]{}
We model the epidemics as stochastic compartmental models where the state of the epidemic is the number of individuals in each compartment. The epidemic evolves probabilistically with events that transition individuals between categories at the rates specified in Fig.\ref{fig.models}, which depend on the state of the epidemic and population size. These models are used as a basis to characterise the optimal policies and various generic features they satisfy.
\begin{tikzpicture}
	\node (SIR) at (40:3cm) [align = center,  draw = externalcircle, fill = innercircle, circle, minimum width = 3cm, line width = 0.15cm] {\textbf{SIR}\\$(S,I) \xrightarrow[]{\mu \frac{SI}{N} h(u)} (S-1,I+1)$\\$(I,R) \xrightarrow[]{\gamma I} (I-1,R+1)$};
	
	\node (SEIR) at (180:3cm) [align = center, draw = externalcircle, fill = innercircle,  circle, minimum width = 3cm, line width = 0.15cm] {\textbf{SEIR}\\$(S,E,I) \xrightarrow[]{\mu \frac{SI}{N} h(u)} (S-1,E+1,I)$\\$(S,E,I) \xrightarrow[]{\epsilon E} (S,E-1,I+1)$\\$(S,E,I) \xrightarrow[]{\gamma I} (S,I-1, R+1)$ };

\node (SI) at (300:4cm) [align = center, draw = externalcircle, fill = innercircle,  circle, minimum width = 3cm, line width = 0.15cm]{\textbf{SI}\\$(S,I) \xrightarrow[]{\mu \frac{I(N-I)}{N} h(u)}(S-1,I+1)$\\$(S,I) \xrightarrow[]{\gamma I} (S,I-1)$};
\end{tikzpicture}
\captionof{figure}{We consider stochastic compartmental models that accurately capture that infections occur randomly in time with controlled rates. }\label{fig.models}
We consider the \acf{SEIR}, \acf{SIR} \cite{Kermack1927,Kendall1956} and \acf{SI} epidemic \cite{Andreson} models for a population of $N$ individuals where individuals may be susceptible $S$, exposed $E$, infected $I$ or recovered $R$. Recoveries occur at a rate proportional to the number of infected individuals $I$ and the parameter $\gamma$. The infection intensity for the \ac{SIR} model is proportional to the average susceptible-infected encounter time $si/N$ and to the transmission rate parameter $\mu$ and lead to individuals transitioning from the susceptible to the infected and exposed category. The exposure events of the \ac{SEIR} epidemic also occur at a rate proportional to the susceptible-infected encounter time and result in individuals transitioning from the susceptible to the exposed category. The mitigation measures $u$ reduce the infection and exposure intensity by $h(u)$ that is a specified yet arbitrary function that responds to changes of the mitigation measure $u$ \cite{Barnett2023}.
The controlled reproduction number\footnote{$\mathcal{R}$ is smaller than or equal to $\mathcal{R}_0$ \cite{Wallinga2004}, irrespectively of the incidence of the disease $S/N$ and hence of herd immunity effects. The latter is accounted by the \emph{effective reproduction number} $\mathcal{R}_e=\mathcal{R}_0h(u)\frac{S}{N}$, \emph{i.e.} a state, intervention and time dependant quantity, which for $S<N$ it satisfies $\mathcal{R}_e\leq\mathcal{R}$. Without any loss of generality we use $\mathcal{R}$ over $\mathcal{R}_e$ in our discussion and analysis to simplify the interpretation of the results. $\mathcal{R}$ is an intervention dependent quantity corresponding to the expected number of cases due to a single infected individual. The controlled $\mathcal{R}$ and $\mathcal{R}_0$ link the parameters of the disease transmission to the possibility of major epidemics \cite{Diekmann1990}. Specifically, $\mathcal{R}>1$ leads to a surges of infections and $\mathcal{R}\leq1$ is associated with an expected decrease of disease prevalence.} of the disease is $\mathcal{R}= \mathcal{R}_0 h(u)$ where $\mathcal{R}_0=\mu/\gamma$ is the basic reproduction number \cite{Dietz1993}, \emph{i.e.} the expected number of new infections due to a single infection.

In our results we consider control policies that are an arbitrary function of the state of the epidemic and in these models we allow for arbitrary epidemic parameters and arbitrary cost parameters for the classes of cost functions we specify.
\end{mabox}

The mitigation measures $u$ can be an arbitrary function of the history of the epidemic and have to be optimised over all possible epidemic outcomes and interventions in order to minimise the average aggregate cost. In particular, the optimal policy $u^*$ achieves the minimum for the problem
\begin{align}\label{eq.GenericCosts}
\min_u\
\mathbb{E}\left[\int_0^\infty g_c(i(t),u(t)) dt\right]
\end{align}
This is a non-trivial problem because the value of $u$ at a certain moment in time influences the future evolution of the epidemic and the interventions at future times. Also, small changes of the costs $g_c(i,u)$ can result in fundamentally different strategies. 

Our approach to this problem is to first develop tools for evaluating exactly the optimal policies over arbitrary feedback policies for general classes of stochastic epidemic models and arbitrary costs $g_c(i,u)$. We then use a combination of analytical (see the Methods section and the Supplementary Information) and computational methods and show that even when arbitrary causal feedback policies are considered, optimal policies for a broad class of cost functions are not sensitive to prevalence making these policies not effective to tame the epidemic.  

\section{Results}\label{sec:results}
We consider the problem in Fig. \ref{fig.MainDiagram} for arbitrary intervention costs that scale with infection levels $g_c(i,u)  = c_1 + c_2 iz(u) + c_3 i$. We find the optimal polices using our computational and analytical tools for arbitrarily complex epidemic models and we observe that irrespectively of the values of the other states the optimal policy is non-increasing in $i$
\begin{align}\label{eq.decreasePolicy}
	u^{\ast}(i + 1) \leq u^{\ast}(i)
\end{align}
Increasing infection levels are not associated with increasingly restrictive mitigation measures and therefore negative feedback polices with respect to prevalence levels are not effective. The optimal policies are also found to be constant with respect to prevalence implying no incentive to adjust the optimally chosen transmission rate to respond to the rapidly changing infection prevalence. After exhaustive computations we find that our results hold for the epidemics in Epidemics models box \ref{modelbox} and for a very broad class of epidemic and cost parameters. We identify these generic features of the optimal policies via our provably exact methods rather than using statistical-learning based methods \cite{Sutton2018} which can yield sub-optimal state dependant policies due to their statistical nature. We also show that \eqref{eq.decreasePolicy} holds analytically for the \ac{SI} epidemic.

The costs apply to influenza-like epidemics \cite{Ludkovski2010} and more broadly to regimes where efforts to control the epidemic are through social distancing, medical isolation, treatment, and quarantine. The proportionality between $i$ and $z(u)$ suggests that cost $c_2 z(u)$ is incurred per individual infected when mitigation measure $u$ is used. The term $c_3 i$ relates to the sole presence of infected individuals and is linked to fatalities and hospitalisations when these are below their capacity. The parameter $c_3$ depends  upon the infected to fatality ratio or the infected to hospitalisation ratio as well as diagnosis efforts.

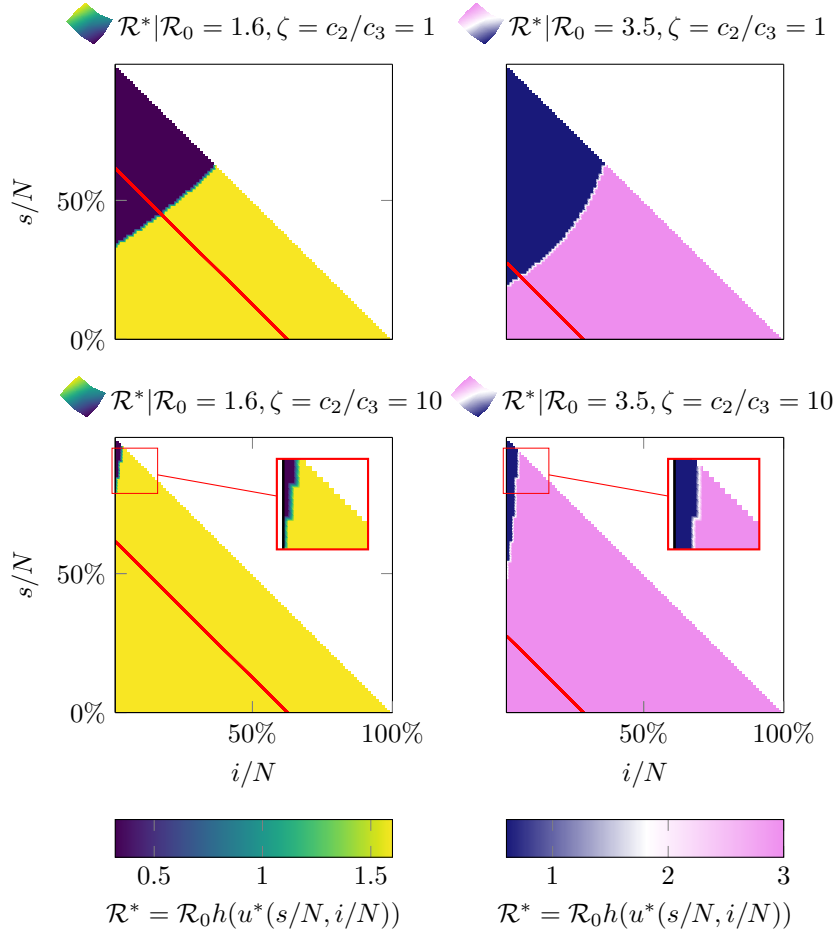
\begin{figure}[h!]
        \centering
\begin{tikzpicture}[spy using outlines={rectangle, magnification=2, connect spies}]
\begin{groupplot}[
    group style             = {group size = 2 by 2, vertical sep = 1.3cm, horizontal sep = 1.5cm, xlabels at = edge bottom , ylabels at = edge left },
    view                    = {0}{90},
    width		    = 0.4\columnwidth,
    height		    = 0.4\columnwidth,
    group/xlabels at 	     = edge bottom,
    ]
\nextgroupplot[ 
	legend style={at={(0.5,+1.05)},anchor=south,draw=none},
    		colormap/viridis,
		unbounded coords=jump,
		ylabel 	            = {$s/N$},
		xticklabel={\pgfmathparse{\tick/1}\pgfmathprintnumber{\pgfmathresult}\%},
		yticklabel={\pgfmathparse{\tick/1}\pgfmathprintnumber{\pgfmathresult}\%},
		xmajorticks = false,
		]
\addplot3[
	surf,
	shader = interp,
	]
	table{SIR_Policies_c2_2pwrgu_2.dat};
\addlegendentry{$\mathcal{R}^*|\mathcal{R}_0=1.6,\zeta = c_2/c_3= 1$}
\addplot3[
        domain=1:(100*(1/1.6)),
        draw                    = red,
	forget plot,
	line cap=rect,
        line width              = 0.03cm,
        style                   = solid,
	](x,{(100*(1/1.6))-x},1.01);\label{plt.HITR1dot6};
\nextgroupplot[ 
	legend style={at={(0.5,+1.05)},anchor=south,draw=none},
    		colormap/violet,
		unbounded coords=jump,
		ymajorticks = false,
		xticklabel={\pgfmathparse{\tick/1}\pgfmathprintnumber{\pgfmathresult}\%},
		yticklabel={\pgfmathparse{\tick/1}\pgfmathprintnumber{\pgfmathresult}\%},
		xmajorticks = false,
		]
\addplot3[
	surf,
	shader = interp,
	]
	table{SIR_Policies3_c2_2pwrgu_2.dat};
\addlegendentry{$\mathcal{R}^*|\mathcal{R}_0=3.5,\zeta = c_2/c_3=1$}
\addplot3[
        domain=1:(100*(1/3.5)),
	forget plot,
        draw                    = red,
	line cap=rect,
        line width              = 0.03cm,
        style                   = solid,
        ](x,{(100*(1/3.5))-x},1.01);
\nextgroupplot[ 
	legend style={at={(0.5,+1.05)},anchor=south,draw=none},
    		colormap/viridis,
		unbounded coords=jump,
		ylabel 	            = {$s/N$},
		xlabel 	            = {$i/N$},
		xticklabel={\pgfmathparse{\tick/1}\pgfmathprintnumber{\pgfmathresult}\%},
		yticklabel={\pgfmathparse{\tick/1}\pgfmathprintnumber{\pgfmathresult}\%},
    		colorbar horizontal, 
		colorbar style = {xlabel={$\mathcal{R}^*=\mathcal{R}_0h(u^*(s/N,i/N))$}},
		every colorbar/.append style={ width= 1*\pgfkeysvalueof{/pgfplots/parent axis height}+0*\pgfkeysvalueof{/pgfplots/group/horizontal sep}},
		]
\addplot3[
	surf,
	shader = interp,
	]
	table{SIR_Policies_c2_3pwrgu_2.dat};
\addlegendentry{$\mathcal{R}^*|\mathcal{R}_0=1.6,\zeta = c_2/c_3= 10$};
\addplot3[
        domain=1:(100*(1/1.6)),
        draw                    = red,
	forget plot,
	line cap=rect,
        line width              = 0.03cm,
        style                   = solid,
	](x,{(100*(1/1.6))-x},1.01);
	\coordinate (spypoint3) at (axis cs:8,87);
        \coordinate (magnifyglass3) at (axis cs:75,75);
        \spy [red, size=1.2cm] on (spypoint3) in node[fill=white] at (magnifyglass3);
\nextgroupplot[ 
	legend style={at={(0.5,+1.05)},anchor=south,draw=none},
    		colormap/violet,
		unbounded coords=jump,
		ymajorticks = false,
		xlabel 	            = {$i/N$},
		xticklabel={\pgfmathparse{\tick/1}\pgfmathprintnumber{\pgfmathresult}\%},
		yticklabel={\pgfmathparse{\tick/1}\pgfmathprintnumber{\pgfmathresult}\%},
    		colorbar horizontal, 
		colorbar style = {xlabel={$\mathcal{R}^*=\mathcal{R}_0h(u^*(s/N,i/N))$}},
		every colorbar/.append style={ width= 1*\pgfkeysvalueof{/pgfplots/parent axis height}+0*\pgfkeysvalueof{/pgfplots/group/horizontal sep}},
		]
\addplot3[
	surf,
	shader = interp,
	]
	table{SIR_Policies3_c2_3pwrgu_2.dat};
\addlegendentry{$\mathcal{R}^*|\mathcal{R}_0=3.5,\zeta = c_2/c_3= 10$};
\addplot3[
        domain=1:(100*(1/3.5)),
        draw                    = red,
	forget plot,
	line cap=rect,
        line width              = 0.03cm,
        style                   = solid,
	](x,{(100*(1/3.5))-x},1.01);
	\coordinate (spypoint4) at (axis cs:8,87);
        \coordinate (magnifyglass4) at (axis cs:75,75);
        \spy [red, size=1.2cm] on (spypoint4) in node[fill=white] at (magnifyglass4);
\end{groupplot}
\end{tikzpicture}
\caption{\textbf{The controlled transmission rate increases as prevalence levels increase.}The diagrams show the optimal $\mathcal{R}^*=\mathcal{R}_0h(u^*(s/N,i/N))$ in four settings for $g_c(i,u)=\zeta iu+i$ corresponding to $\zeta=c_2/c_3=\{1,10\}$ and $\mathcal{R}_0=\{1.6,3.5\}$. The controlled reproduction number $\mathcal{R}^*$ is constant for low prevalence levels and is non-decreasing for increasing prevalence, \emph{i.e.} the optimal policies are predominantly constant and are non-increasing for increasing prevalence levels. Negative feedback schemes using infection prevalence are therefore not effective for taming the epidemic and there is no incentive to vary interventions. The diagrams display the values of $\mathcal{R}^*$ in the lower left triangle, \emph{i.e.} the sub diagonals of the graph, and correspond to regimes where recoveries have taken place. That is the values of $s/N$ and $i/N$ such that $s/N+i/N < 1$ and $r/N>0$. The top-left to bottom-right main diagonal of the diagrams corresponds to regimes where $s/N+i/N =1$, \emph{i.e.} regimes where there are no recovered individuals and the \ac{SIR} and \ac{SI} model have an exact correspondence. Note that $H_i$ (\ref{plt.HITR1dot6}) is a threshold on the recovered levels such that $s/N+i/N=1-H_i$ and is thus shown as a diagonal.
Negative feedback schemes on the infection levels are not effective irrespective of the specific value of $s$ in that the optimal $\mathcal{R}^\ast$ is non-decreasing irrespective of $H_i$ and of the value of $s$. For $\zeta=1$, and for prevalence up to approximately $50\%$, it is optimal to implement interventions also beyond $H_i$. For $\zeta=10$ it is optimal to rely upon herd immunity since $\mathcal{R}^*$ increases to $\mathcal{R}_0$ before $H_i$ is reached.}\label{fig.PoliciesSIRSummary}
\end{figure}

To illustrate the practical significance of our findings we show in Fig. \ref{fig.PoliciesSIRSummary} four representative cases for the \ac{SIR} epidemic. 
The optimal $\mathcal{R}^*$ for any incidence level $s/N$ does not change with prevalence levels and is non-decreasing returning to the  value of the basic reproduction number $\mathcal{R}_0$ at a level that depends on the costs.  Surprisingly, these constant policies outperform negative feedback policies with respect to prevalence levels in controlling the epidemic irrespectively of the relative cost coefficient $\zeta=c_2/c_3$ and basic reproduction number $\mathcal{R}_0$. The search for the optimal policy can therefore be restricted to that for a constant policy and the more complex dynamic and data-driven surveillance based threshold policies bring no benefit. Prevalence measurements, that are the focus of many studies \cite{PrevStudy22}, remain however necessary in order to estimate key transmission parameters, such as $\mathcal{R}_0$ and $h(u)$, that are also required to obtain the policy. If prevalence is not directly observed and needs to be inferred then there is still no incentive to use negative feedback as the corresponding policy would be in this case even less effective in reducing the cost.
\newpage

\subsection{Herd immunity}
The uncontrolled transmission rate depends also upon the number of susceptible $s$. Specifically, low values of $s$ reduce the transmission rate via \emph{herd immunity} \cite{Metcalf2020} a phenomenon that unfolds at slow timescales. In the optimal policies however negative feedback at the faster timescales of the infection dynamics is not observed irrespectively of the levels of $s$. In particular, depending on the parameters, there is no incentive to vary the constant optimal $\mathcal{R}^*$ up to the \emph{herd immunity threshold} \cite{Fine2011,Fontanet2020} $H_i$
\begin{equation}\label{eq.HIT}
	H_i = (1-1/\mathcal{R}_0)
\end{equation}
The threshold $H_i$ (see (\ref{plt.HITR1dot6}) in Fig. \ref{fig.PoliciesSIRSummary}) corresponds to the proportion of population that has to be removed from the susceptible pool via recoveries or vaccinations \cite{Delamater2019,Diekmann1990} so that, in the absence of interventions, the expected prevalence decreases\footnote{Analogously the effective reproduction number $\mathcal{R}_e=1$.}.

For low values of intervention-to-hospitalization cost ratio $\zeta = c_2/c_3$ , interventions can be beneficial even after $H_i$ since these can reduce the transmission of the disease at low cost. In these regimes, which are associated with low values of $s$, interventions are however expected to be reduced in practice due to the indirect suppression of transmission from herd immunity effects. For high values of intervention-to-hospitalization cost ratio it is optimal to rely upon herd immunity effects rather than intervene to try to control the disease.

\subsection{Total costs are higher for negative feedback schemes}
We display in Fig. \ref{fig.CostsSIRfullObserv}, as an illustrative example, the ratio between the optimal total cost and the total costs associated to the following ad-hoc chosen  sub-optimal negative feedback policy \eqref{eq.suboptimalu}
\begin{equation}\label{eq.suboptimalu}
	u_{fb}(i) =
	\begin{cases}
		0 \text{ if } i\leq i_{th},\\
		u_{N_u}, \text{ otherwise }
	\end{cases}
\end{equation}
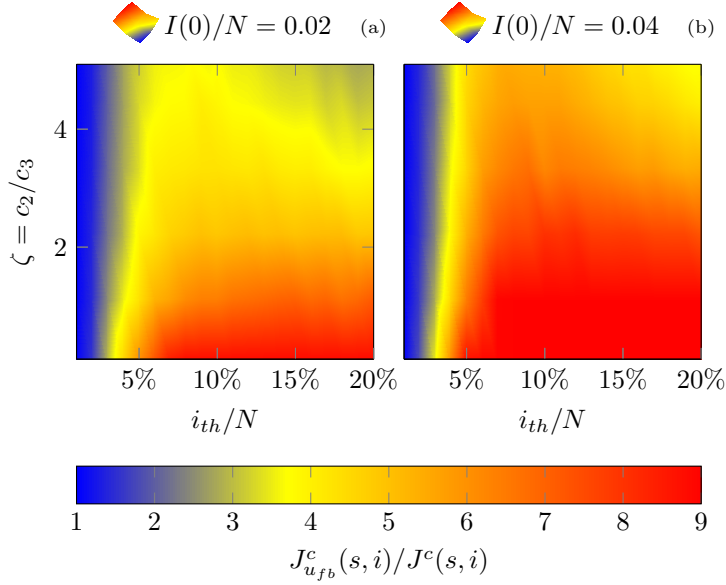
\begin{figure}[h!]
        \centering
\begin{tikzpicture}
\begin{groupplot}[
    group style             = {group size = 2 by 1, vertical sep = 0.9cm, horizontal sep = 0.4cm, xlabels at = edge bottom , ylabels at = edge left },
    view                    = {0}{90},
    width		    = 0.42\columnwidth,
    height		    = 0.42\columnwidth,
    ymin                    = 0.1,
    ymax                    = 5.1,
    group/xlabels at 	     = edge bottom,
    group/ylabels at  	     = edge left,
    ]
\nextgroupplot[
		legend style={at={(0.5,+1.05)},anchor=south,draw=none},
		ylabel 	            = {$\zeta = c_2/c_3$},
		xlabel 	            = {$i_{th}/N$},
		xticklabel={\pgfmathparse{\tick/1}\pgfmathprintnumber{\pgfmathresult}\%},
    		point meta min          = 1,
    		point meta max          = 9,
    		colorbar horizontal, 
		colorbar style = {xlabel={$J^c_{u_{fb}}(s,i)/J^c(s,i)$}},
		every colorbar/.append style={ width= 2*\pgfkeysvalueof{/pgfplots/parent axis height}+1*\pgfkeysvalueof{/pgfplots/group/horizontal sep}},
		]
\addplot3[
	surf,
	shader = interp,
	]
	table{SIR_RatioCosts2pc.dat};
\addlegendentry{$I(0)/N=0.02$};
\nextgroupplot[ 
	legend style={at={(0.5,+1.05)},anchor=south,draw=none},
    		point meta min          = 1,
    		point meta max          = 4,
		xticklabel={\pgfmathparse{\tick/1}\pgfmathprintnumber{\pgfmathresult}\%},
		ymajorticks = false,
		xlabel 	            = {$i_{th}/N$},
		]
\addplot3[
	surf,
	shader = interp,
	]
	table{SIR_RatioCosts4pc.dat};
\addlegendentry{$I(0)/N=0.04$};
\end{groupplot}
\node[text width=6cm, align=center, anchor = south] at (group c1r1.north east) {\subcaption{\label{fig_costs_sir_2pc}}}; 
\node[text width=6cm, align=center, anchor = south] at (group c2r1.north east) {\subcaption{\label{fig_costs_sir_4pc}}};
\end{tikzpicture}
\caption{\textbf{The total costs of using negative feedback policies prevalence are high.} The diagrams show the ratio between the total expected costs $J^c_{u_{fb}}$ associated with the sub optimal prevalence sensitive feedback policy of the type in \eqref{eq.suboptimalu} over the expected costs $J^c$ associated with the optimal policies that are not sensitive to prevalence levels for which negative feedback is not effective. The total costs of adopting a negative feedback scheme on the infections is several times higher than for the optimal case. The initial conditions are set to $I(0)/N = 0.02$ and $I(0)/N=0.04$. The results are shown for $\gamma=0.32, \mathcal{R}_0=3.5$.}\label{fig.CostsSIRfullObserv}
\end{figure}
Irrespectively of cost parameters and the initial conditions, using a negative feedback scheme on the infection yields higher costs relative to the implementation of the optimal constant mitigation measures. The diagrams also show that the costs of sub optimal negative feedback schemes that vary interventions at low values of $i_{th}$ are close to the optimal costs. This is because for low values of $i_{th}$ the sub optimal policies are comparable to the optimal policies and result in short periods of exponential growth of the infections before the interventions are used.

\subsection{Effectiveness of negative feedback}
Mitigating epidemics via negative feedback with respect to prevalence can be effective for costs functions that include targeted and population wide costs\footnote{It should also be noted that the considered costs can be formulated in the form $g_c(i,u)=c 1 + c 2 iu + c 3 i + c 4 (N - i)u$, that frequently appear in the literature \cite{Rowthorn2020a,Barnett2023}} $g_c(i, u) = c_1 + c_2 iu + c_3 i +c_4Nu$ where $c_4$ is the cost coefficient for measures that are applied to  the entire population $N$. These are associated with stay-at-home strategies, travel restrictions or business closure orders applied to the entire population. These mitigation measures are required for high case-to-fatality ratio and for regimes where stay-at-home interventions are costly compared with the targeted interventions and even with respect to hospitalisation costs $c_3$.
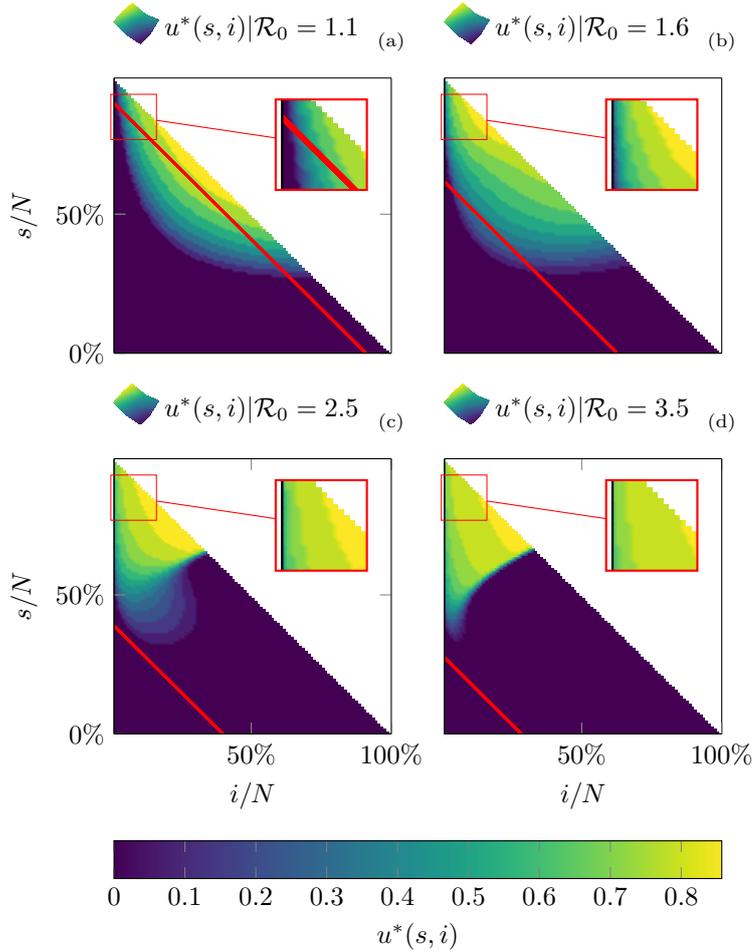
\begin{figure}[h!]
        \centering
\begin{tikzpicture}[spy using outlines={rectangle, magnification=2, connect spies}]
\begin{groupplot}[
    group style             = {group size = 2 by 2, vertical sep = 1.4cm, horizontal sep = 0.7cm, xlabels at = edge bottom , ylabels at = edge left },
    view                    = {0}{90},
    width		    = 0.4\columnwidth,
    height		    = 0.4\columnwidth,
    group/xlabels at 	     = edge bottom,
    group/ylabels at  	     = edge left,
    ]
\nextgroupplot[
		legend style={at={(0.45,+1.1)},anchor=south,draw=none},
    		colormap/viridis,
		ylabel 	            = {$s/N$},
		unbounded coords=jump,
		xticklabel={\pgfmathparse{\tick/1}\pgfmathprintnumber{\pgfmathresult}\%},
		yticklabel={\pgfmathparse{\tick/1}\pgfmathprintnumber{\pgfmathresult}\%},
		xmajorticks = false,
		]
\addplot3[
	surf,
	shader = interp,
	]
	table{SIR_Policies_additiveforFBR01.dat};
\addlegendentry{$u^*(s,i)|\mathcal{R}_0 = 1.1$};
\addplot3[
        domain=1:(100*(1/1.1)),
        draw                    = red,
        forget plot,
        line cap=rect,
        line width              = 0.03cm,
        style                   = solid,
        ](x,{(100*(1/1.1))-x},1.01);\label{plt.HITRed}
\coordinate (spypoint1) at (axis cs:8,85);
        \coordinate (magnifyglass1) at (axis cs:75,75);
        \spy [red, size=1.2cm] on (spypoint1) in node[fill=white] at (magnifyglass1);
\nextgroupplot[ 
	legend style={at={(0.45,+1.1)},anchor=south,draw=none},
    		colormap/viridis,
		unbounded coords=jump,
		ymajorticks = false,
		xmajorticks = false,
		]
\addplot3[
	surf,
	shader = interp,
	]
	table{SIR_Policies_additiveforFBR02.dat};
\addlegendentry{$u^*(s,i)|\mathcal{R}_0 = 1.6$};
\coordinate (spypoint2) at (axis cs:8,85);
        \coordinate (magnifyglass2) at (axis cs:75,75);
        \spy [red, size=1.2cm] on (spypoint2) in node[fill=white] at (magnifyglass2);
\addplot3[
        domain=1:(100/1.6),
        draw                    = red,
        forget plot,
        line cap=rect,
        line width              = 0.03cm,
        style                   = solid,
        ](x,{100*(1/1.6)-x},1.01);
\nextgroupplot[ 
	legend style={at={(0.45,+1.1)},anchor=south,draw=none},
    		colormap/viridis,
		unbounded coords=jump,
		ylabel 	            = {$s/N$},
		xlabel 	            = {$i/N$},
		xticklabel={\pgfmathparse{\tick/1}\pgfmathprintnumber{\pgfmathresult}\%},
		yticklabel={\pgfmathparse{\tick/1}\pgfmathprintnumber{\pgfmathresult}\%},
    		colorbar horizontal, 
		colorbar style = {xlabel={$u^*(s,i)$}},
		every colorbar/.append style={ width= 2*\pgfkeysvalueof{/pgfplots/parent axis height}+1*\pgfkeysvalueof{/pgfplots/group/horizontal sep}},
		]
\addplot3[
	surf,
	shader = interp,
	]
	table{SIR_Policies_additiveforFBR03.dat};
\addlegendentry{$u^*(s,i)|\mathcal{R}_0 = 2.5$};
\coordinate (spypoint3) at (axis cs:8,85);
        \coordinate (magnifyglass3) at (axis cs:75,75);
        \spy [red, size=1.2cm] on (spypoint3) in node[fill=white] at (magnifyglass3);
\addplot3[
        domain=1:(100/2.5),
        draw                    = red,
        forget plot,
        line cap=rect,
        line width              = 0.03cm,
        style                   = solid,
        ](x,{100*(1/2.5)-x},1.01);
\nextgroupplot[ 
	legend style={at={(0.45,+1.1)},anchor=south,draw=none},
    		colormap/viridis,
		unbounded coords=jump,
		xlabel 	            = {$i/N$},
		xticklabel={\pgfmathparse{\tick/1}\pgfmathprintnumber{\pgfmathresult}\%},
		ymajorticks = false,
		]
\addplot3[
	surf,
	shader = interp,
	]
	table{SIR_Policies_additiveforFBR04.dat};
\addlegendentry{$u^*(s,i)|\mathcal{R}_0 = 3.5$};
\coordinate (spypoint4) at (axis cs:8,85);
        \coordinate (magnifyglass4) at (axis cs:75,75);
        \spy [red, size=1.2cm] on (spypoint4) in node[fill=white] at (magnifyglass4);
\addplot3[
        domain=1:(100/3.5),
        draw                    = red,
        forget plot,
        line cap=rect,
        line width              = 0.03cm,
        style                   = solid,
        ](x,{100*(1/3.5)-x},1.01);
\end{groupplot}
\node[text width=6cm, align=center, anchor = south] at (group c1r1.north east) {\subcaption{\label{fig_SIR_FB_1}}}; 
\node[text width=6cm, align=center, anchor = south] at (group c2r1.north east) {\subcaption{\label{fig_SIR_FB_2}}};
\node[text width=6cm, align=center, anchor = south] at (group c1r2.north east) {\subcaption{\label{fig_SIR_FB_3}}};
\node[text width=6cm, align=center, anchor = south] at (group c2r2.north east) {\subcaption{\label{fig_SIR_FB_4}}};
\end{tikzpicture}
\caption{\textbf{Negative feedback policies can be effective for population wide costs.} The diagrams display the optimal policies for the \ac{SIR} epidemic with costs $g_c(i,u)=c_2 iu+c_3i+c_4Nu, c_2 \approx 0 \text{ and } \upsilon =c_4/c_3=50$. In all cases in an optimal policy for increasing levels of prevalence increasingly restrictive interventions are to be implemented. Interventions become more restrictive as prevalence increases: a result in agreement with recent studies \cite{Barnett2023} where parametric uncertainty is also considered. For moderate reproduction numbers (see Fig. \ref{fig_SIR_FB_1} and \ref{fig_SIR_FB_2}) interventions are gradually enforced after an outbreak. The interventions are adopted progressively as infection levels grow to approximately $i/N\approx 35\%$. For higher values of $\mathcal{R}_0$, (see Fig.\ref{fig_SIR_FB_3} and \ref{fig_SIR_FB_4}) despite the high costs it is optimal to adopt some degree of interventions even for very low values of prevalence. The strictest mitigation measures are then to be adopted progressively for increasing levels of prevalence. (\ref{plt.HITRed}) indicates the herd immunity threshold $H_i$ for the considered $\mathcal{R}_0$. The optimal policies change irrespectively of $H_i$. }\label{fig.PoliciesSIRFB_paper}
\end{figure}
For low values of quarantine and isolation costs compared to population wide measures we find that prevalence sensitive policies can be effective to mitigate an epidemic. We show this in the representative Fig. \ref{fig.PoliciesSIRFB_paper} where taming the spread incurs high costs even for few infections and thus there is little incentive to make use of interventions for low prevalence as these apply to the entire population. Interventions are therefore likely after a certain infection level is reached or progressively for increasing prevalence. Identifying cost parameters has therefore a fundamental effect upon effectively controlling epidemics and our findings show that efforts should be focused towards identifying costs and epidemic parameters as these will help identify the right measures, rather than always aiming to vary the measures with changes in prevalence.

\end{bibunit}

\section{Methods} \label{sec:Methods}
\renewcommand{\refname}{Methods References}
\begin{bibunit}[naturemag]
We describe in this section the methods used to derive our results, with the Supplementary Information providing further details. To facilitate readability, the results and the corresponding methodology are summarized in Theory box~\ref{theorybox}, with the text in Section \ref{sec:Methods}  providing a more detailed description.

We develop a unified framework for computing the optimal policies in arbitrarily complex stochastic epidemic models formulated as \ac{MJP} encompassing a broad class of practically relevant models \cite{Kraemer2025,Giordano2020}.
We provide a method based on uniformisation, a natural way to convert the considered problem to that of an exact discrete time decision process enabling the computation of the policies via the Bellman equation and the value iteration algorithm. Unlike methods commonly used in the context of reinforcement learning that rely on approximate time discretisations or deterministic differential models (both of which introduce uncontrolled errors) our framework preserves the stochastic dynamics and yields provably exact policies.

\begin{pabox}[label={theorybox},breakable]{}
Finding explicitly an optimal policy is in general a non-trivial problem, which is further complicated by the stochastic and continuous time nature of the evolution of epidemics. It requires optimising over an indefinite amount of time and over arbitrary functions of the history of the process in order to minimise costs $g_c(x,u)$ by modulating the transition probabilities $p^u_{x,y}$.
Here $x$ is the current number of individuals in each of the considered compartments and $y$ is the number of individuals in the considered compartments after the transition has occurred (see Section \ref{meth.SysJumps} and Supplementary Information).

Addressing this problem in discrete time lies in satisfying the Bellman equation \cite{Bellman1957,Bellman1952}, a necessary condition that optimal policies have to satisfy. The Bellman equation is an equilibrium relation for the total costs $J(x)= \min_u[g_c(x,u) + \sum_j p^{u^{*}}_{x,y}J(y)]$, quantifying that deviations from the optimal policy inevitably yield higher total costs.

We develop computational tools based upon equivalent discrete time formulations, for finding the optimal policies for arbitrarily complex continuous time stochastic epidemic models. We use these to deduce the features stated in the main text (Section \ref{sec:results}) for the optimal policies for the \ac{SEIR} and \ac{SIR} models. For the \ac{SI} model we also prove this analytically for arbitrary epidemic and costs parameters via induction arguments.

Specifically, in order to reduce the complexity of searching over an infinite space we associate the continuous time control problem to that of an appropriately constructed equivalent \emph{uniformised}  problem  (see Fig. \ref{fig.MCUnif}, Section \ref{meth.Control} and Supplementary Information). The uniformised problem is an equivalent discrete time problem with a reduced computational complexity over continuous time formulations requiring to optimise only over a finite set of parameters. The resulting problem has an \emph{exact} correspondence with the original continuous time problem, in the sense that the resulting optimal policy is the same for both problems and can be found using the Bellman equation.

Since the epidemics are considered over an indefinite period of time the optimal policies are also independent of time \cite{Bertsekas2005} and depend only on the current state of epidemic further simplifying the search of the policy.

We explicitly compute the optimal policies for the \ac{SEIR} and \ac{SIR} epidemic using an iteration based algorithm, the \emph{value iteration algorithm} \cite{Howard2007}, that converges to the optimal policies which satisfy the Bellman equation (See Section \ref{meth.DP} and Supplementary Information). We then use this iteration based algorithm to deduce the features of the optimal control policies and observe that using feedback prevalence signals is not effective to modulate control measures. We use this algorithm also to prove analytically for the \ac{SI} epidemic (see Section \ref{meth.SI} and Supplementary information) that for arbitrary parameters and arbitrary policies the optimal policies do not incorporate negative feedback with respect to infections. The analytical derivations use the structure of the Bellman equation to carry out various induction arguments revealing an incentive to decrease or at most hold constant the measures for increasing infections, \emph{i.e.} $u^*(i+1)\leq u^*(i)$.

\end{pabox}

\subsection{Epidemic modeling as Markov Jump Processes}\label{meth.SysJumps}
We model the evolution of the epidemic as a continuous-time \ac{MJP}, where a closed population of $N$ individuals transition between compartments via random events occurring at the rates specified below. The models we consider quantify more accurately the evolution of the epidemic compared to approximate deterministic models since the disease progression is inherently stochastic and only the infection and recovery rates can be controlled.

	The random state of the epidemic $X(t)$ at time $t$ is the number of individuals in the considered compartments (\emph{i.e.} Susceptible, Exposed, Infected and Recovered). We denote a particular configuration that the epidemic can assume as $x$. The epidemic evolves via infection, exposure and recovery events with individuals transitioning between the compartments summarised below in Table \ref{tab.Systems}. Each event $i$ is associated with a rate $W_i(x,u)$ and a change of the state \emph{e.g.} $(S,I,R)\to(S-1,I+1,R)$. The rates at which these events occur depends on the state of the epidemic $x$ and the control input $u$ that modulates the transmission through a function $h(u)$ (see Section 2 of the Supplementary Information for more details).
\begin{table}[h!]
        \centering
        \begin{tabular}{ccc}%
            \toprule
            \textbf{Event} & \textbf{State transition} & \textbf{Rate}\\
                    \midrule
	    \ac{SEIR} Exposure & $(S,E,I,R) \xrightarrow[]{} (S-1,E+1,I,R)$ & $\frac{\mu}{N} SI\cdot h(u) $\\
	    \ac{SEIR} Infection & $(S,E,I,R) \xrightarrow[]{} (S,E-1,I+1,R)$ & $\epsilon E $\\
	    \ac{SEIR} Recovery & $(S,E,I,R) \xrightarrow[]{} (S,E,I-1, R+1)$  & $\gamma I$\\
	    \ac{SIR} Infection & \emph{$(S,I,R) \xrightarrow[]{} (S-1,I+1,R)$} & $\frac{\mu}{N} SI\cdot h(u)$\\
	    \ac{SIR} Recovery & \emph{$(S,I,R) \xrightarrow[]{} (S,I-1,R+1)$}   &  $\gamma I$\\
	    \ac{SI} Infection & $I \xrightarrow[]{} I+1$ & $\frac{\mu}{N} I(N-I)h(u)$\\
      	    \ac{SI} Recovery & $I \xrightarrow[]{} I-1$  & $\gamma I$\\
                \bottomrule
        \end{tabular}
	\caption{Events associated with the \acf{SEIR}, \acf{SIR} and \acf{SI} epidemics. The controlled infection and exposure events are proportional to the probability of individuals encountering while recovery rates depend linearly on infection levels.}\label{tab.Systems}
\end{table}

The epidemic takes a particular state for a time that is exponentially distributed with parameter $\xi(x, u)=\sum_k W_k(x,u)$ that corresponds to the total rate at which any of the $k$ leaving events may occur at state $x$ under control $u$.

We illustrate our methods on \ac{SEIR}-like models (see \emph{e.g.} Fig. \ref{fig.MCUnif}) but our results apply to any compartmental epidemic process described by a continuous-time \ac{MJP} with arbitrary number of compartments and complexity over finite state space with state-dependent rates (see \emph{e.g.} \cite{Giordano2020}). Our methods can also account for heterogeneous interactions (see Supplementary Information).

We next seek to compute the optimal policies that select $u$ dynamically to minimise cumulative costs while accounting for the system's inherent randomness.
\subsection{Control problem and uniformisation}\label{meth.Control}
We consider control strategies that modulate the disease transmission in real time to achieve epidemic control, while minimizing costs $g_c(x,u)$ associated with interventions and infections, over an infinite time horizon, i.e.
	\begin{equation}\label{eq.OCP}
	u^*(x) = \argmin_{u}\lim_{T\to +\infty} \mathbb{E}\left[\int_0^T g_c(X(t),u) dt\middle\vert X(0)=x_0\right]
\end{equation}
subject to the epidemic state vector $X$ evolving according to one of the epidemics described in the previous section. The problem is a continuous time \ac{MDP} \cite{Puterman1994}, where the control $u$ modulates the transition rates between states in order to minimise the total expected costs. The solution to this problem corresponds to finding an optimal policy $u^*(x)$ that minimises a prescribed socio-economic cost $g_c(x,u)$ related to the state of the epidemic. In particular, $g_c$ associates for each infected level a cost that society has to incur and a costs for interventions that reduce the transmission of the disease (see Section 3 of Supplementary Information for more details).

In the optimal control problem \eqref{eq.OCP} the state of the system and the control input $u$ change at random moments in time $T_k$ and are otherwise constant. It is sufficient therefore to evaluate the policy only when state transitions take place and reduce the problem to a probabilistically equivalent discrete time one for which the optimal policy is the same to the one in problem \eqref{eq.OCP}. Analogously, the cost $g_c$ is also constant between events and can be evaluated at the transition times $T_k$. This method, known as uniformisation, enables the exact computation of the optimal policies while avoiding the approximations inherent to time discretisation techniques that fail to capture these dynamics precisely.

\begin{figure}[!h]
	\centering
\scalebox{0.56}{
\begin{tikzpicture}[
	node distance=2.8cm,
	color = red,
	->,
	>=latex,
	auto, 
	every edge/.append style={thick,line width=0.45mm, color={rgb,255:red,40; green,42; blue,115}},  
    state/.style={
        circle,
        draw={rgb,255:red,21; green,45; blue,80},   
        fill={rgb,255:red,174; green,197; blue,230},
        text=black,
        minimum size=1.5cm,
        line width=0.5mm
    	}
	]
  \begin{scope}
  \node[state] (s01) {\shortstack{$s=0$\\$i=1$}};  
  \node[state] (s00) [left  of = s01]{\shortstack{$s=0$\\$i=0$}};  
  \node[state] (s10) [above of=s00] {\shortstack{$s=1$\\$i=0$}};  
  \node[state] (s20) [above of=s10] {\shortstack{$s=2$\\$i=0$}};  
  \node[state] (s02) [right of=s01] {\shortstack{$s=0$\\$i=2$}};  
  \node[state] (s03) [right of=s02] {\shortstack{$s=0$\\$i=3$}};  
  \node[state] (s11) [above of=s01] {\shortstack{$s=1$\\$i=1$}};  
  \node[state] (s12) [right of=s11] {\shortstack{$s=1$\\$i=2$}};  
  \node[state] (s21) [above of=s11] {\shortstack{$s=2$\\$i=1$}};  
  \node[state] (s30) [above of=s20] {\shortstack{$s=3$\\$i=0$}};  

  \path (s02) edge[bend left] node{$\gamma 2$}     (s01)
	(s03) edge[bend left] node{$\gamma 3$}     (s02)
	(s12) edge[bend left] node{$\gamma 2$}     (s11)
	(s11) edge[bend left] node{$\gamma 1$}     (s10)
	(s21) edge[bend left] node{$\gamma 1$}     (s20)
	(s01) edge[bend left] node{$\gamma 1$}     (s00)
	(s11) edge[bend right,color={rgb,255:red,127; green,20; blue,22}] node{$\mu / N h(u)$}     (s02)
	(s21) edge[bend left,color={rgb,255:red,127; green,20; blue,22}] node{$\mu / N2 h(u)$}     (s12)
	(s12) edge[bend left,color={rgb,255:red,127; green,20; blue,22}] node{$\mu / N2 h(u)$}     (s03);
\node [above of =s30]{\parbox{0.3\linewidth}{\subcaption{}\label{subfig_rates}}};
  \end{scope}
  
  \begin{scope}[xshift=11cm]
  \node[state] (s01) {\shortstack{$s=0$\\$i=1$}};  
  \node[state] (s00) [left  of = s01]{\shortstack{$s=0$\\$i=0$}};  
  \node[state] (s10) [above of=s00] {\shortstack{$s=1$\\$i=0$}};  
  \node[state] (s20) [above of=s10] {\shortstack{$s=2$\\$i=0$}};  
  \node[state] (s02) [right of=s01] {\shortstack{$s=0$\\$i=2$}};  
  \node[state] (s03) [right of=s02] {\shortstack{$s=0$\\$i=3$}};  
  \node[state] (s11) [above of=s01] {\shortstack{$s=1$\\$i=1$}};  
  \node[state] (s12) [right of=s11] {\shortstack{$s=1$\\$i=2$}};  
  \node[state] (s21) [above of=s11] {\shortstack{$s=2$\\$i=1$}};  
  \node[state] (s30) [above of=s20] {\shortstack{$s=3$\\$i=0$}};  
\node [above of =s30]{\parbox{0.3\linewidth}{\subcaption{}\label{subfig_unif}}};

  \path (s02) edge[bend left] node{$\frac{\gamma 2}{\nu}$}     (s01)
	(s03) edge[bend left] node{$\frac{\gamma 3}{\nu}$}     (s02)
	(s12) edge[bend left] node{$\frac{\gamma 2}{\nu}$}     (s11)
	(s11) edge[bend left] node{$\frac{\gamma 1}{\nu}$}     (s10)
	(s21) edge[bend left] node{$\frac{\gamma 1}{\nu}$}     (s20)
	(s01) edge[bend left] node{$\frac{\gamma 1}{\nu}$}     (s00)
	(s11) edge[bend right,color={rgb,255:red,127; green,20; blue,22}] node{$\frac{\mu / N h(u)}{\nu}$}     (s02)
	(s21) edge[bend left,color={rgb,255:red,127; green,20; blue,22}] node{$\frac{\mu / N2 h(u)}{\nu}$}     (s12)
	(s12) edge[bend left,color={rgb,255:red,127; green,20; blue,22}] node{$\frac{\mu / N2 h(u)}{\nu}$}     (s03)
	(s21) edge [in=130,out=175,loop,color={rgb,255:red,214; green,118;blue,48}] node {$1-\frac{\mu 2/N h(u)+\gamma 2}{\nu}$} (s21)
	(s12) edge [in=50,out=80,loop,color={rgb,255:red,214; green,118;blue,48}] node {$1-\frac{\mu 2/N h(u)+\gamma 2}{\nu}$} (s12)
	(s03) edge [in=50,out=80,loop,color={rgb,255:red,214; green,118;blue,48} ] node {$1-\frac{\gamma 3}{\nu}$} (s03)
	(s02) edge [in=50,out=80,loop,color={rgb,255:red,214; green,118;blue,48} ] node {$1-\frac{\gamma 2}{\nu}$} (s02)
	(s01) edge [in=130,out=175,loop,color={rgb,255:red,214; green,118;blue,48}] node {$1-\frac{\gamma 1}{\nu}$} (s01)
    	(s11) edge [in=130,out=175,loop,color={rgb,255:red,214; green,118;blue,48}] node {$1-\frac{\mu/N h(u)+\gamma 2}{\nu}$} (s11);
  \end{scope}
\end{tikzpicture}
}\caption{\textbf{An example of the uniformisation method for \ac{SIR} epidemic.} The figure shows in Fig. \ref{subfig_rates} the states of the \ac{SIR} \ac{MJP} and in Fig. \ref{subfig_unif} the associated uniformised version for a population of $N=3$ individuals. The diagram \ref{subfig_rates} shows the events and the associated rates for the \ac{SIR} epidemic for the susceptible and infected individuals. The recovered category does not appear since the population is assumed constant and it follows that $R=N-S-I$ and the state $s=3, i=0$ corresponds to a non infected population with no events leading to it or leaving from it. The diagram includes infection events (\protect\tikz[baseline=-0.5ex]\protect\draw[myinfectcolor,thick](0,0)--(0.5,0);)
with susceptible individuals becoming infected at the rate $\mu\frac{si}{N} h(u)$, and recovery events (\protect\tikz[baseline=-0.5ex]\protect\draw[myrecoverycolor,thick](0,0)--(0.5,0);) with infected individuals recovering at the rate of $\gamma i$. The digram \ref{subfig_unif} shows the uniformised version of the \ac{SIR} epidemic where in addition to the events of the original process model the fictitious self events (\protect\tikz[baseline=-0.5ex]\protect\draw[myorange,thick](0,0)--(0.5,0);) leave the state unchanged and ensure the expected time in each state is equal to that of the original process.}\label{fig.MCUnif} \end{figure}
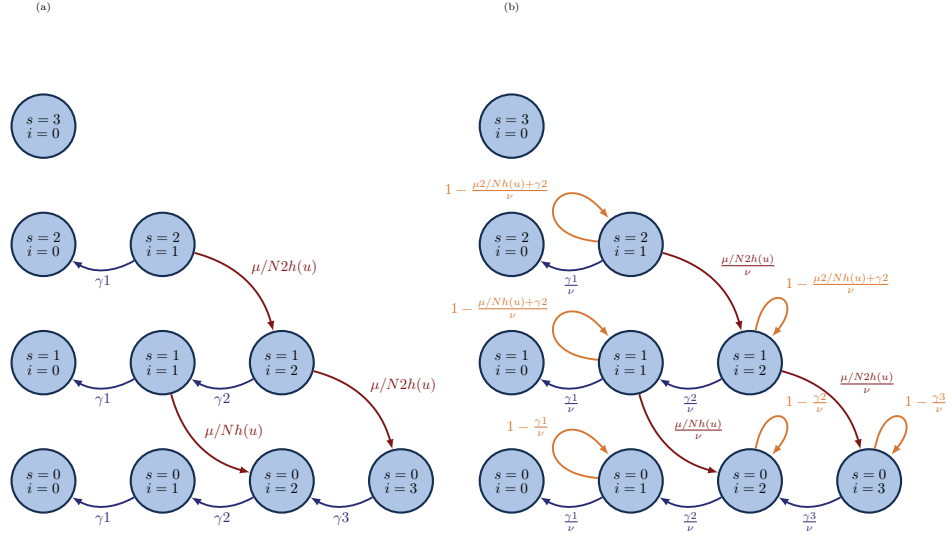

The key to our transformation is the uniformisation rate $\nu$, that is chosen to be greater than or equal to the inter-event time parameter $\xi(x,u)$ in any state
\begin{equation}\label{eq.Unifpar}
	\nu \geq \max_{x,u} \xi(x,u)
\end{equation}
This choice makes the events of the resulting uniformised process occur at a constant rate $\nu$. In order to obtain a probabilistically equivalent problem we introduce appropriate fictitious self events that leave the state unchanged (see Fig. \ref{fig.MCUnif}). These events ensure that the process is in each state an equivalent amount of time to that of the continuous time version (see Section 5 of Supplementary Information for more details).
Self events transitions occur at a higher rate, but these leave the state unchanged ensuring the expected time in each state is equal to the original process. This is done by accounting for the difference between the uniform rate and the actual rate of change of the original process \emph{i.e.} $\nu -\xi(x,t)$.

The probability $p^u_{x,y}$ that the uniformised process transitions from state $x$ to state $y$ are obtained as follows. For the $k$ outgoing events that transition the state of the epidemic from $x$ to a different state $y$ the transition probabilities are $p^u_{x,y}=W_k(x,u)/\nu$ \emph{i.e.} the ratio of the specific transition rate of the original process $W_k(x,u)$ from $x$ to $y$ over the faster rate $\nu$. The self event at state $x$ that accounts for the excess of probability flow due to the increased rate $\nu$ has transition probability $p^u_{x,x}= 1- \xi(x,u)/\nu$.

The instantaneous costs $g_c(x,u)$ for the optimal control problem \eqref{eq.OCP} are also transformed to account for the changes at the random moments in time $T_k$. Since events occur at rate $\nu$ scaling the instantaneous cost by $1/\nu$ yields the same expected cost per unit time as in the original problem. This reduces the original continuous time formulation to a discrete time \ac{MDP} formulation with instantaneous costs $g=g_c(x,u)/\nu$ for which the optimal policy is an optimal policy also for \eqref{eq.OCP}.

Our method allows to apply dynamic programming techniques, that are otherwise difficult to apply in continuous time and guarantees exact optimal policies for epidemic models with arbitrary number of compartments, which would otherwise require approximations. The full derivation and conditions under which uniformisation guarantees exactness are provided in the Supplementary Information.

\subsection{Dynamic Programming and Policy Computation}\label{meth.DP}
In order to determine the optimal policies $u^*$ we formulate the uniformised version of continuous time problem in \eqref{eq.OCP} discussed in the previous section. Here the state $x$ is the number of individuals in the considered compartments and the optimal action has to be chosen from the finite set of available interventions $\mathcal{U}$. 

To explicitly obtain the optimal policies we make use of the Bellman equation for optimality that characterises the minimum expected costs $J(x)$ starting from any state $x$ as the sum of intermediate costs and future costs. This recursive relation follows from the dynamic programming principle \cite{Bellman1954,Bellman1952} and quantifies that deviations from the optimal policy inevitably yield higher total costs
\begin{equation}\label{eq.bellman}
	J(x)= \min_{u\in\mathcal{U}}[g(x,u) + \sum_{y\neq x} p^{u}_{x,y}J(y)]
\end{equation}
where $g(x,u)$ is the instantaneous cost of applying measure $u$ in state $x$ and $p^u_{x,y}$ is the transition probability (as derived in the previous section) from state $x$ to $y$ under intervention $u$.

In order to compute the optimal policies that satisfy \eqref{eq.bellman} we make use of the value iteration algorithm. This is an iterative method that computes the total expected cost
\begin{equation}\label{eq.vi}
J_{k+1}(x)= \min_{u\in\mathcal{U}}[g(x,u) + \sum_{y\neq x} p^{u^{*}}_{x,y}J_k(y)]
\end{equation}
where $J_k(x)$ is the value function at iteration $k$ and $J_0(x)$ is an arbitrary guess for the initial conditions. The algorithm iterates over all possible states, updating the value function $J_k$. Once the value function $J_k(x)$ has converged, the optimal policy $u^*$ for any state $x$ is then computed by selecting the action $u^*(x)$ that minimises \eqref{eq.bellman}. More detailed convergence properties and additional implementation details are provided in the Supplementary Information Section 5.

\subsection{Analytical Results for Specific Models}\label{meth.SI}
We show analytically that for the \ac{SI} epidemic models the optimal policy $u^\ast$ satisfies \eqref{eq.decreasePolicy}, \emph{i.e.} increasing infection levels are not associated with increasingly restrictive mitigation measures. The result is shown to hold for arbitrary cost and epidemic parameters for costs of the form $g_c(i,u)  = c_1 + c_2 iz(u) + c_3 i$. The derivation of the results is based on induction arguments on the value iteration algorithm, which allows to ultimately show that negative feedback is not effective for epidemic control. Detailed derivations and further examples of optimal policy structures are provided in Sections 3 and 5 of the Supplementary Information.

\subsection{Computational studies parameters}\label{ss.parameters}
The value of the parameters considered for Fig.\ref{fig.PoliciesSIRSummary}, Fig. \ref{fig.CostsSIRfullObserv} and Fig. \ref{fig.PoliciesSIRFB_paper} are representative of COVID-19 and flu epidemics. These are selected using various studies, detailed below, including data from the European Centre of Disease Prevention and Control. These are only indicative figures and we provide more complete analysis in the Supplementary Information where we obtain similar results for a large range of functions and parameters that characterise a broad set of epidemics. It should also be noted that for the \ac{SI} epidemic we have characterised the optimal policy for arbitrary parameters and transmission rate reduction functions $h$.

The value of $\gamma$ is held at $\gamma = 0.32$ \cite{Lin2020}. The values of $\mathcal{R}_0=3.5$ is in line with the values found for COVID-19 in \cite{Flaxman2020,Liu2020,Korolev2021} and \cite{ArroyoMarioli2021} and $\mathcal{R}_0=1.6$ can be associated with Ebola \cite{Driessche2017} or the 2009 H1N1 flu pandemic \cite{Biggerstaff2014,Petersen2020}. The specific form of the optimal policies depends also upon the specific $h(u)$. Assessing $h(u)$ is however challenging especially for novel pathogens \cite{Ferguson2020} and suffers from high uncertainty \cite{Ferguson2007}. Frequently considered settings are linear $h(u)=(1-q_{eff}u)$ \cite{Alvarez2021} and quadratic $h(u)=(1-q_{eff}u)^2$ \cite{Barnett2023}. The parameter $q_{eff}\in [0,1]$ captures the effectiveness of contact tracing and of the interventions \cite{Eames2003}. Linear transmission rate control functions $h(u)=(1-q_{eff}u)$ elicit rapid decreases of the measures while quadratic $h(u)=(1-q_{eff}u)^2$ yield more gradual changes (see Supplementary Information). Prevalence signals are ineffective to modulate interventions also for low $q_{eff}$ where interventions are insufficient to achieve $\mathcal{R}^* \leq 1$ irrespectively of the specific transmission rate control function $h(u)$. Policy makers however may seek to suppress in a best effort way the spread of the disease or reduce very high peaks of infection curves for example when pharmaceutical interventions and vaccines are in development.

\backmatter

\bmhead{Data availability}
The results are based on simulated data and all figures have been generated using a combination of the Latex Tikz package and the pgfplots package.

\bmhead{Code availability}
The code used for the computation of the policies and the simulation of the epidemics is made available via the Zenodo platform with instructions for replicating the analysis at \url{https://zenodo.org/records/15733511} 

\bmhead{Competing Interests} The authors do not have any competing interests to declare.

\bmhead{Supplementary information}
Supplementary Information is available for this paper. The Supplementary Information comprises of: 1 Introduction; 2 Preliminaries; 3 Problem formulation and considered costs; 4 Results; 5 Derivation of the results; Supplementary Appendix A detailing the derivation of the results and Appendix B includes additional figures and discussion.

\end{bibunit}

{\includepdf[pages=-, pagecommand={\thispagestyle{empty}}]{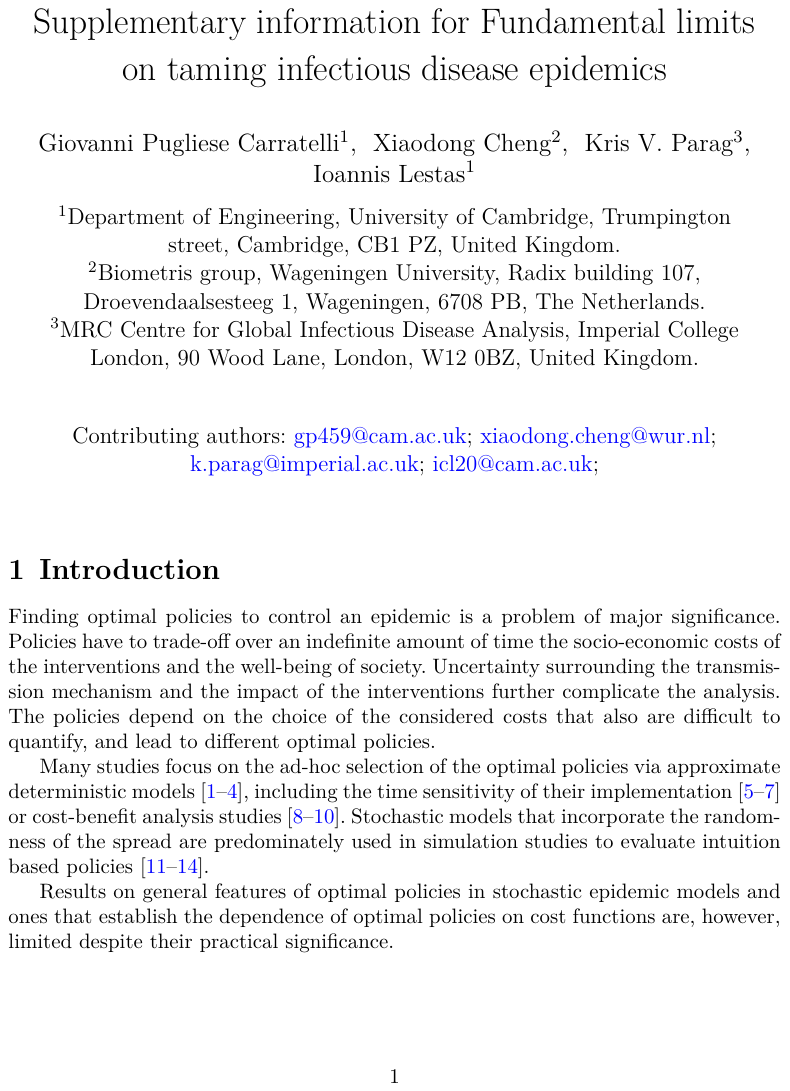}}
\end{document}